\documentclass[12pt]{iopart}

\usepackage{graphicx}
\usepackage{amssymb}
\usepackage{color}
\usepackage{verbatim}
\usepackage{graphicx}
\usepackage{epsfig}
\usepackage{bm}
\usepackage[ragged]{sidecap}
\usepackage{dsfont}
\usepackage{ulem}

\begin{document}

\title{Underlining some limitations of the statistical formalism in quantum mechanics: \\
Reply to the Comment of Bodor and Di\'osi}

\author{F Fratini$^{1}$ and A G Hayrapetyan$^{2}$}
\address{$^1$
Department of Physics, Post Office Box 3000, FI-90014, University of Oulu, Finland}
\ead{fratini@physi.uni-heidelberg.de}
\vspace{0.2cm}
\address{$^2$
Max-Planck-Institut f\"ur Kernphysik, Postfach 103980, 69029 Heidelberg, Germany}

\begin{abstract}
In a paper of us, it is showed that Density Matrices do {\it not} provide a complete description of 
ensembles of states in quantum mechanics, since they lack measurable information concerning the preparation of the ensembles. Bodor and Di\'osi have 
later posted a comment on that article, which agrees on some points of it but disagrees on some others.
This reply is intended to clarify the discussion.
\end{abstract}

\pagestyle{empty}

In a paper of us, it is showed that density matrices do {\it not} provide a complete description of 
ensembles of states in quantum mechanics, since they lack measurable information concerning the preparation of the ensembles \cite{FraHyr2011}.
We exposed this lack by theoretically analyzing the Variance of the spin along a chosen direction, for two differently prepared ensembles of states.
The Variance obtained by analyzing one ensemble turned out to be different from the Variance obtained by analyzing the other ensemble. On the other hand, the density matrices
of both ensembles turned out to be the same. Thus, clearly it follows that the quantum mechanical description of an ensemble of states given by its
density matrix does not encompass all the measurable\footnote{The Variance can be easily measured in experiments.} characteristics the ensemble possesses.

\medskip

On this, Bodor and Di\'osi (BD) have more recently posted a comment in arXiv arguing that i) our analysis is irrelevant
for standard statistical ensembles\footnote{
By ``statistical ensemble'' is meant an ensemble whose populations of states are statistically determined by means of a certain statistical distribution which guarantees random mixing.}, ii) our conclusions about the limitations of standard theory are unjustified \cite{BD2011}.
While we agree on the first point above (though we think it is irrelevant to the analysis carried out in \cite{FraHyr2011}), we disagree on the second. \\
The analysis we carried out in \cite{FraHyr2011} does not include nor mention
statistical ensembles, as we explicitly define and refer only to ``prepared ensembles''\footnote{By ``prepared ensembles'' we mean ensembles whose
populations of states are fully determined.}. In our work, in fact, the two ensembles $\mathcal{A}$ and $\mathcal{B}$ 
have {\it exactly} the same amount of particles with spin defined along and opposite to a chosen direction. The chosen direction
is different for the two ensembles. These two ensembles are definitely not statistical ensembles.
Yet their description in terms of density matrices, as it will more extensively explained further on, is to be considered pertaining to the statistical formalism of Quantum Mechanics \cite{Sak, Blum}.\\
For the same reasons, we do not furthermore agree on the BD's sentence ``Fratini and Hayrapetyan are apparently unaware of [the fact that] their ensembles are different from what standard statistics as well as standard quantum mechanics understand as statistical ensembles.
Statistical ensembles must consist of independent states, this requires random mixing.''

\medskip

BD then lay their comment out on statistical ensembles. They state several times that they agree on our findings, but our conclusions are irrelevant since the ensembles taken in \cite{FraHyr2011} should have not been taken, as they are not statistical ensembles. Instead, we should have
taken, in their opinion, statistical ensembles to write the article out. Our reply to this statements
is just that we considered prepared ensembles, because we are basically free to decide which case study to investigate. 

\medskip

From what above and from BD's arguing, we may summarize that BD's complains are mostly about the misuse of the word ``statistical'' in Ref.~\cite{FraHyr2011}.
In their opinion, since statistical ensembles have not been considered, the word ``statistical'' may not be used.
However, as discussed in \cite{FraHyr2011}, prepared ensembles, as well as any other kind of ensemble, are ensembles of particles with different states. Because of this, the outcome of an experiment on the ensemble is reasonably postulated to be the {\it statistical average} of the outcomes on the states the ensemble is made of. It is this postulate that permits the definition of ``density operator'', also called ``statistical operator'' \cite{Bal}, for the description of the ensemble. \\
Our opinion is that the word ``statistical'' is pertinent to the quantum mechanical description of any ensemble as long as the {\it statistical average} of the outcomes is postulated. 
Of the same opinion are Sakurai \cite{SakTr}, Fano \cite{Fano} and Greiner \cite{Greiner}.
The expressions 
``density operator'' and ``statistical operator'', as well as the expressions ``density matrix'' and ``statistical matrix'', are in fact commonly interchangeable in Quantum Mechanics \cite{Bal, Fano}.\\
In conclusion, Statistical Quantum Mechanics is commonly understood to deal with the description of ensembles of states, irrespective of as to whether or not they are statistical, prepared or whatever else kind of ensembles.

\medskip

On their statement ``Had Fratini and Hayrapetyan used the correct mixing to construct the two ensembles they would have left with no measurable difference between them which fact is fundamental in the quantum theory and is particularly well understood in quantum informatics'',
we fully agree. ``Correct mixing'' must be here understood as ``correct mixing for obtaining correct statistical ensembles''.

\medskip

The two-particle density matrices of our ensembles $\mathcal{A}$ and $\mathcal{B}$ are different, as BD noticed. 
Nonetheless, one must remember that one-particle density matrices are the kind of density matrices which are widely and almost uniquely used in literature. Many books, articles and theorems are written for one-particle density matrices. On the other hand, two- or many-particle density matrices are considered very seldom and only for specific purposes. The intent of \cite{FraHyr2011} has been also to show that information concerning the preparation of an ensemble, while it is invisible from the point of view of the one-particle density matrix of the ensemble, can be highlighted in experiments. 

\medskip

The final discussion in BD's article \cite{BD2011}, not contained in the first version of the manuscript \cite{BD2011_1}, is about a possible attempt to calculate the Variance within the density matrix formalism in Quantum Mechanics. They observe that the $N$-particle density matrix must be considered for that purpose, not the one- neither the two-particle density matrices. $N$ is here the number of particles the ensemble contains. We do agree on this issue too, though nothing has been said on how to do that. To this regard, we must also consider the fact that, as showed in section ``Discussion'' of Ref. \cite{FraHyr2011}, a definition of Variance as quantum mechanical operator is problematic and leads to contradictions. Thus, a fully quantum mechanical calculation of the Variance will not be easy nor immediate, in either case we use one- or many-particle density matrices\footnote{In Ref. \cite{FraHyr2011}, the Variance has been computed by taking the quantum mechanical prediction on the single-particles measurements and by then applying Classical Statistics.}.\\
We want here also to stress the fact that the density matrix formalism has been introduced into Quantum Mechanics precisely to avoid the long (but most correct) writing of $N$ quantum states.
If the $N$-particle density matrix must be invoked in order to correctly calculate the Variance of an ensemble of $N$ states, then the density matrix description loses evidently its value and becomes useless. In realistic ensembles, $N$ is of the order of the Avogadro Number ($\sim 10^{23}$), so that neither writing nor dealing with its $N$-particle density matrix is clearly feasible. This underlines, as well as our article does, the limitations of the statistical formalism of Quantum Mechanics, i.e. of the density matrix (or density operator) description of ensembles of states, when dealing with quantities such as the Variance.



\newpage


\section*{References}

\begin{thebibliography}{21}

\bibitem{FraHyr2011}	F. Fratini and A. G. Hayrapetyan, Phys. Scr. {\bf 84}, 035008 (2011), arXiv: 1108.6249 .
\bibitem{BD2011}			Andr\'as Bodor and Lajos Di\'osi, arXiv:1110.4549v2
\bibitem{Sak}					J.J. Sakurai, {\it Modern Quantum Mechanics} (Addison-Wesley, 1994).
\bibitem{Blum}				K. Blum, {\it Density Matrix Theory and Applications} (NewYork Plenum, 1996).
\bibitem{Bal}					V. V. Balashov, A. N. Grum-Grzhimailo and N. M. Kabachnik, 
											{\it Polarization and Correlation Phenomena in Atomic Collisions} (New-York Kluwer, 2000).
\bibitem{SakTr}				Ref. \cite{Sak},  pg. 182.
\bibitem{Fano}				U. Fano, {\it Description of States in Quantum Mechanics by Density Matrix and Operator Techniques}, 
											Rev. Mod. Physics {\bf 29}, 74 (1957).
\bibitem{Greiner}			W. Greiner, {\it Thermodynamics and Statistical Mechanics} (Springer, 1995), pg. 255.
\bibitem{BD2011_1}		Andr\'as Bodor and Lajos Di\'osi, arXiv:1110.4549v1



\end{thebibliography}
\end{document}